\documentclass[12pt,preprint]{aastex}

\def\p/{\mbox{$^1$}}
\def\pp/{\mbox{$^2$}}
\def\ppp/{\mbox{$^3$}}
\def\pppp/{\mbox{$^4$}}
\def\m/{\mbox{$^{-1}$}}
\def\mm/{\mbox{$^{-2}$}}
\def\mmm/{\mbox{$^{-3}$}}
\def\mmmm/{\mbox{$^{-4}$}}
\def\Ms/{\mbox{M$_\odot$}}
\def\ebv{\mbox{$E(4405-5495)$}}
\def\rv{\mbox{$R_{5495}$}}

\shorttitle{Progenitor of SN 2004dj}
\shortauthors{Ma\'{\i}z-Apell\'aniz et al.}

\slugcomment{Submitted to Astrophysical Journal Letters}

\begin{document}

\title{The Progenitor of the Type II-P SN 2004dj in NGC 2403}

\author{J. Ma\'{\i}z-Apell\'aniz\altaffilmark{1,2}, 
        Howard E. Bond\altaffilmark{3}, and M. H. Siegel\altaffilmark{4}}
\affil{Space Telescope Science Institute\altaffilmark{5}, 3700 San Martin 
Drive, Baltimore, MD 21218, U.S.A.}

\author{Y. Lipkin, D. Maoz, E. O. Ofek, and D. Poznanski}
\affil{School of Physics and Astronomy, Tel-Aviv University, Tel-Aviv 69978,
Israel}



\altaffiltext{1}{Affiliated with the Space Telescope Division of the European 
Space Agency, ESTEC, Noordwijk, Netherlands.}
\altaffiltext{2}{e-mail contact: {\tt jmaiz@stsci.edu}.}
\altaffiltext{3}{Visiting Astronomer, Kitt Peak National Observatory, National
Optical Astronomy Observatory, which is operated by the Association of
Universities for Research in Astronomy, Inc., under cooperative agreement with
the National Science Foundation.}
\altaffiltext{4}{Current address: Department of Astronomy, University of Texas,
Austin, TX 78712.}
\altaffiltext{5}{The Space Telescope Science Institute is operated by the
Association of Universities for Research in Astronomy, Inc., under NASA
contract No. NAS5-26555.}

\begin{abstract}

The Type II-P supernova 2004dj in the nearby galaxy NGC~2403 occurred at a
position coincident with object 96 in the list of luminous stars and clusters
in this galaxy published by Sandage in 1984. The coincidence is established
definitively through astrometric registration of our ground-based archival
images of NGC~2403 with our recent images showing the SN\null. The archival
images show that Sandage~96 is slightly resolved from the ground. Pre-outburst
blue spectrograms obtained by Humphreys and Aaronson reveal that
Sandage~96 has a composite spectrum, dominated in the blue region by A- and
B-type stars, while infrared photometry shows that Sandage~96 also contains red
supergiants. These results demonstrate that Sandage~96 is a young compact
cluster.  We have studied the stellar population of Sandage~96, using published
photometric measurements combined with a chi-square-fitting code. We derive a
cluster age of 13.6~Myr, a reddening of $\ebv=0.172$, and a total stellar mass
of 24,000~$M_\odot$\null. For this cluster age, the SN progenitor had a
main-sequence mass of $15\,M_\odot$. Post-outburst photometry of Sandage~96 may
establish whether the progenitor was a red or blue supergiant.

\end{abstract}

\keywords{galaxies: individual (NGC 2403) --- galaxies: star clusters ---
          supernovae: general --- supernovae: individual (SN 2004dj)}

\section{Introduction}

Type II supernovae (SNe) are believed to arise from core collapses of evolved
massive stars, yet there is surprisingly little direct evidence in
the form of pre-outburst observations of progenitor stars to support this
belief. To date, the best-observed SN progenitor is the blue supergiant
precursor of SN~1987A in the Large Magellanic Cloud, but the pre-outburst data
are limited to photographic and photoelectric photometry and a photographic
objective-prism spectrum \citep[see][]{Walbetal89}. More recently, the
progenitors of (or binary companions of) two further Type II SNe have
been identified on archival images from the {\it Hubble Space Telescope\/}
({\it HST\/}) and ground-based telescopes \citep{Maunetal04,Smaretal04}.
See \citet{Smaretal03} for further discussion of the importance of efforts
to identify SN progenitors.

The bright SN 2004dj was discovered in the nearby spiral galaxy NGC 2403 on
2004 July 31 by K.~Itagaki \citep[see][]{Naka04}, and it was quickly classified
spectroscopically as a normal Type II-P (plateau) core-collapse SN caught about
3 weeks after the explosion \citep{Pataetal04}.   NGC 2403 is a member of the
M81 group, with a distance of about 3.3~Mpc \citep{Karaetal04}; SN~2004dj is
thus the nearest SN to be discovered since SN~1993J in M81, and it is destined
to be a  well-observed SN~II-P event.   Type II-P SNe have been reviewed by
\citet{Leonetal02} and \citet{Hamu03}; Hamuy's Table~1, which lists the
properties of 24 well-documented SN~II-P events, shows that SN~2004dj is the
nearest Type~II-P SN yet discovered (apart from the peculiar Type~II-P
SN~1987A)\null.

In this {\it Letter}, we use recent images of the SN along with archival images
to show that SN~2004dj coincides with a previously cataloged object in
NGC~2403, which proves to be a slightly resolved young cluster that contains
dozens of massive stars.  Well-calibrated photometry and digital spectrograms 
of this cluster exist in the literature. Using models of starburst populations,
we are able to estimate the age, mass, and other properties of this compact
cluster, and to set limits on the nature of the progenitor star.

\section{Identification of the Progenitor}

\subsection{Sandage 96 in NGC 2403}

\citet{Sand84} conducted a photographic survey of the brightest blue and red
supergiants in NGC~2403. Based on the reported position of SN~2004dj, several
authors \citep[see][]{Yamaetal04} have pointed out the close coincidence
of the SN with a luminous blue object in Sandage's list, cataloged as no.~96. 
This object is designated ``Sandage~96'' hereafter. 

Sandage 96 would be one of the brightest blue supergiants in NGC~2403, if it
were a single star.  However, \citet{Sand84} annotated it as a possible
cluster. \citet{Hump80} obtained a blue spectrum of this object and classified
it as B5:~I:, but suggested a composite nature. An improved blue spectrum was
obtained by \citet{HumpAaro87}, and is reproduced in their Figure~1. They
report that the integrated spectrum resembles that of an early A-type star, but
that the hydrogen Balmer lines are too broad to be due solely to an A
supergiant, and the presence of \ion{He}{1} lines shows that there are
also B-type stars present in the spectrum. Humphreys \& Aaronson also did not
detect H$\alpha$ emission. They concluded that Sandage~96 is a compact cluster,
without an accompanying \ion{H}{2} region. 

Sandage~96 was included as object n2403-3866 in a list of young massive
clusters in nearby galaxies by \citet{Lars99}, who measured
Johnson-Kron-Cousins $UBVI$ photometry. The object is also present in the 2MASS
infrared point-source catalog \citep{Skruetal97}, from which we obtained its
$JHK_{\rm s}$ photometry. The optical and infrared photometric data from these
sources are given in Table~\ref{photometry}, where we have removed the
reddening correction applied by Larsen to his measurements. The data definitely
do not correspond to any single star, since the color is blue at short
wavelengths but an infrared excess begins to set in around the $I$ band and is
pronounced in the infrared.

\subsection{Archival and New Images}

The site of SN 2004dj was imaged (serendipitously) by H.E.B. on three nights in
1999 January with the Mosaic CCD camera on the Mayall 4-m telescope at Kitt
Peak National Observatory (KPNO).  All of the observations were made with
standard Johnson-Kron-Cousins $B$, $V$, and $I$ filters, along with a
Thuan-Gunn $u$ filter. 



Following the discovery announcement of SN~2004dj, CCD images of the SN were
obtained by Y.L., D.M., E.O.O., and D.P., using the 1.0-m telescope of the Wise
Observatory and $UBVRI$ filters.  Figure~1 (left) shows the field of the SN and
a few neighboring bright field stars, as imaged in the $V$ band with the Wise
telescope on 2004 August~4.  In Figure~1 (right) we show the same field, as
extracted from an $I$-band KPNO Mosaic frame taken on 1999 January~19.  In
order to verify the association of the SN with Sandage~96 which is strongly
suggested by Fig.~1, we combined four of the best Wise Observatory CCD frames
of SN 2004dj  to produce a fairly deep image.  We then registered this frame
with the Kitt Peak 4-m frame shown in Fig.~1, using 8 nearby field stars to
determine the geometric transformation of the Wise frame onto the Kitt Peak
image.  The rms of the astrometric fit is only $0\farcs03$ in each coordinate,
in spite of the $2\farcs6$ seeing of the Wise image and the fact that several
of the reference stars are saturated in the KPNO image.  We find that SN~2004dj
coincides with Sandage~96 to within $0\farcs07$ ($0\farcs04$ in RA, $0\farcs06$
in Dec).  

Sandage~96 is slightly non-stellar on our archival frames.  In our $I$ frame
with the best seeing, $0\farcs8$, Sandage~96 has a FWHM of $1\farcs0$. 
S.~Smartt (2004, private communication) has informed us that  archival images
taken with the 8.2-m Subaru telescope in 0\farcs44 seeing through an H$\alpha$
filter (which just measures continuum flux as there is no line emission) show
a FWHM for Sandage~96 of 0\farcs6. 

It should be noted that, at the distance of  NGC~2403, $1\arcsec$ corresponds
to a linear scale of 16~pc and that 4~pc is a typical size for a compact young
stellar cluster (see, e.g. \citealt{Maiz01b}). There is thus no doubt that the
SN lies well within the bounds of the cluster Sandage~96, and that it must have
arisen from a star belonging to the cluster.\footnote{After this letter was
submitted, {\it HST\/} observations of SN~2004dj were obtained on 2004 August
16 and 17. Based on long-exposure  ACS/WFC images, \citet{FiliLi04} report (and
we confirm independently) that, although the SN is heavily saturated in these
frames, it definitely took place at a position coincident with
Sandage~96. We have analyzed higher-resolution ACS/HRC images that were also
obtained and find that: (a)~in the short-exposure (non-saturated) images only a
very bright point-source (the SN itself) is present, which is so bright that it
makes detection of any other stars in the cluster extremely difficult; and
(b)~in the NUV-optical objective-prism exposures the SN is easily detected, as
well as three nearby (within a few arcseconds) stars also detected in the
ground-based KPNO image. All of the {\it HST\/} images confirm the coincidence
between SN 2004dj and  Sandage~96, but it will take additional observations,
after the SN has subsided, to determine the color-magnitude diagram and other
properties of the cluster.} 


\section{The Stellar Population of Sandage 96}

\subsection{Mathematical Approach}

In order to determine the nature of the stellar population within the compact
cluster and thus constrain the main-sequence mass and other properties of the
SN~2004dj progenitor, we used CHORIZOS \citep{Maiz04c}, a
chi-square-minimization code that finds which members of a family of spectral
energy distributions (SEDs) are compatible with the observed integrated colors
of a stellar population. CHORIZOS allows the user to select different input SED
families and extinction laws, and to place statistical constraints on the
fitted parameters. 

Although the above discussion shows that Sandage~96 is a cluster, for
completeness we tested whether this object could have been a single luminous
supergiant or had to be a compact cluster, by using two different SED families,
one consisting of single stars, and the other of cluster populations.   For the
stellar models we selected \citet{Kuru04} atmospheres with low gravities, solar
metallicity, and effective temperatures between 3,500 K and 50,000 K\null. For
the cluster models, we selected Starburst99 \citep{Leitetal99} model
populations with solar metallicity, a Salpeter IMF, an upper-mass cutoff of
$100 \, M_\odot$, and ages between $10^6$ and $10^{10}$~yr. The starburst models
assume that all of the stars in the population were created at the same time.
The choice of solar metallicities for the models is justified by the 
galactocentric distance of Sandage 96 \citep{Fieretal86}.

CHORIZOS was executed using both SED families, selecting a \citet{Cardetal89}
interstellar extinction law with $\rv = 3.1$, and two free parameters: the
extinction, \ebv, and either the effective temperature (for the stellar models)
or log(age) (for the cluster models).\footnote{\ebv\ and $R_{5495}$ are the
monochromatic equivalents to $E(B-V)$ and $R_V$, respectively. Here  4405 and
5495 are the assumed central wavelengths (in \AA) of  the $B$ and $V$ filters,
respectively. CHORIZOS uses monochromatic quantities because $E(B-V)$ and $R_V$
depend  not only on the amount and type of dust but also on the SEDs.} Since we
are fitting two parameters and using six colors (derived from seven
magnitudes), the problem has 4 degrees of freedom.

\subsection{Results of the Population Fitting}

Our results from CHORIZOS decisively eliminate the possibility that Sandage~96
is a single luminous star. The best fit for the Kurucz models is
for a highly reddened O~star, but its reduced $\chi^2$ is
43, indicating the extremely poor quality of the fit.

However, cluster models provide an excellent fit, with a best
reduced $\chi^2$ of 0.28 (which, if anything, suggests that photometric
uncertainties may have been overestimated). The likelihood map produced by
CHORIZOS (Fig.~\ref{likelihood}) shows two peaks in the log(age)-\ebv\ plane,
indicating the existence of two solutions compatible with the available
photometry. Properties for both solutions are shown in
Table~\ref{results}. The ``young'' solution (age of 13.6~Myr) is the one that
has the highest likelihood. The ``old'' solution (age around 29~Myr) is less
likely, but its validity cannot be immediately rejected at the 10\% level,
since its reduced $\chi^2$ is 1.74. The SED for the ``young''
solution is shown in  Fig.~\ref{spectra}. 

Fig.~2 shows that the fit to the ``young'' solution is excellent. For the
``old'' solution, the largest contribution to $\chi^2$ is from the
$U$ band, because the Balmer jump of the SED is large compared
to
the observed photometry (and to the spectrum shown by  \citet{HumpAaro87}.
Therefore, we strongly prefer the ``young'' solution.

This solution yields a reddening of $\ebv=0.172\pm0.022$,
which
is in very good agreement with
the $E(B-V) = 0.18$ obtained by \citet{Pataetal04} from the \ion{Na}{1} D
equivalent
width in the SN spectrum.

The turnoff mass for a cluster at an age of 13.6~Myr is $15\,M_\odot$, and the
corresponding main-sequence spectral type is B1~V.

\section{Discussion}

The age of 13.6~Myr derived above is also compatible with other
secondary evidence. The blue spectrum of a 13.6 Myr old cluster is dominated by
blue giants and supergiants, so its classification as composite B5: I: by
\citet{Hump80} agrees with our expectation. Furthermore, NGC~2403 is a galaxy
that has undergone recent intense episodes of star formation that have produced
several massive young clusters and OB associations
\citep{Drisetal99b,Maiz01b}. Some of those clusters are only a few  Myr old and
show strong H$\alpha$ emission; however, as noted above, Sandage~96 is not
detected in H$\alpha$ \citep[see also][]{Sivaetal90}. This is what is expected
for a 13.6 Myr old cluster, since such an object would have had enough time to
disperse its parent  molecular cloud by stellar winds and supernova explosions,
and should have no O stars left [apart from, possibly, O stars formed as a
result of mass transfer in binaries \citep{Cerv98}]. 

From the known distance and measured $V$ magnitude, we can determine the total
mass of the stellar population\footnote{The mass estimate assumes a Salpeter 
IMF between $1 M_\odot$ and $100 M_\odot$.}. The result, 
$\approx$24,000$\,M_\odot$, makes Sandage~96 intermediate between the most
massive young  clusters in NGC~2403 and typical Galactic open clusters, with a
mass similar to those of some of the ``rich'' young clusters in the LMC\null.
Therefore, Sandage~96 is likely not massive enough to survive for a Hubble time
and become a future globular cluster; rather it is destined to dissolve into
the general field of NGC~2403 \citep{FallZhan01}.

The uncertainties quoted in Table~\ref{results} should be taken with caution,
since they do not include external error sources, such as uncertainties in the
distance, stellar models, and other assumptions, including the assumed
instantaneous starburst episode. In particular, we should note that current
stellar evolutionary models have problems in producing the right numbers and
types of red supergiants in the 6-30 Myr age interval
\citep{LangMaed95,MassOlse03}. Therefore, it is possible that the actual age
for Sandage~96 may differ from our value by 1 or 2~Myr. Unless the cluster is
very compact, future {\it HST\/}  imaging should provide a color-magnitude
diagram that will clarify these issues.

Another point is that our starburst models assume an infinite number of stars,
whereas the actual number of bright stars in Sandage~96 must be relatively low,
leading to a potential problem of stochastic sampling. However, as we show in
Table~\ref{results}, the predicted numbers of red supergiants in Sandage~96 
(the main contributors to the NIR SED) and of blue giants and supergiants (the
main contributors to the blue-visible spectrum) are above the critical value of
10 \citep{CervVall03} below which sampling effects start to become significant.

Supernovae of Type II-P are expected to have red supergiant progenitors with
initial masses lower than those of Type II-L and Ib/c \citep{Hegeetal03}. The
turnoff mass we find for Sandage~96, $15\,M_\odot$, is consistent with that of
other Type II-P SNe \citep{Smaretal03}. However, it should be pointed out that
the heretofore best-studied SN progenitor of a core-collapse supernova, that of
SN~1987A, turned out to be a blue supergiant. As seen in Table~\ref{results},
our models predict that $\approx$12 red and 2-3 blue supergiants should have
been present in Sandage~96 prior to the explosion of SN~2004dj.   Given that
red supergiants are the dominant sources of the NIR flux in a 13.6~Myr old
cluster,  and that blue giants and supergiants dominate in the $U$ and $B$
bands, we can  make a prediction that will test the nature of the progenitor.
Once the SN has faded away, if the star that exploded was a red supergiant,
then the NIR integrated magnitudes of the cluster should have dimmed by about
0.08~mag, while the blue magnitudes will be essentially unchanged. If, on the
other hand, the progenitor was a blue supergiant,  the $UBV$ photometry should
be dimmer by $\approx$0.04~mag. However, it will take several years before the
SN has faded enough to make a negligible contribution to the total flux of
Sandage~96 \citep{Pozzetal04,Zhanetal04}.

We conclude that SN 2004dj took place in Sandage~96, an $\approx$14 Myr old  
compact cluster of intermediate mass. Based on its SN type of II-P and the
$15\,M_\odot$ turnoff mass of the cluster, the progenitor is likely to have
been a red supergiant, of which $12\pm4 $ existed in Sandage~96 prior to the
explosion. Although a blue-supergiant progenitor cannot be ruled out completely
at this stage, our population models suggest that a comparison of pre- and
post-outburst photometry of Sandage~96 may determine whether the exploded
progenitor was blue or red. 

\begin{acknowledgements}

We would like to thank Nolan Walborn for helpful discussions; Leonardo \'Ubeda
for help with evolutionary tracks; and the referee, Stephen Smartt, for very
helpful suggestions and for communicating results in advance of publication.
H.E.B. acknowledges support from the NASA
UV, Visible, and Gravitational Astrophysics Research and
Analysis Program through grant NAG5-3912. D.M. acknowledges support by the
Israel Science Foundation---the Jack Adler Foundation for Space Research, Grant
63/01-1.

\end{acknowledgements}

\bibliographystyle{apj}
\bibliography{general}

\begin{deluxetable}{cr}
\tablecaption{Published Photometry}
\tablewidth{0pt}
\tablehead{\colhead{Quantity} & \colhead{Value}}
\startdata
$V$         &   18.05\phn\phn$\pm$\phn0.03\phn \\
$U-B$       & $-$0.48\phn\phn$\pm$\phn0.06\phn \\
$B-V$       &    0.30\phn\phn$\pm$\phn0.03\phn \\
$V-I$       &    0.87\phn\phn$\pm$\phn0.03\phn \\
$J$         &      16.193\phn$\pm$\phn0.081    \\
$H$         &      15.539\phn$\pm$\phn0.106    \\
$K_{\rm s}$ &      15.416\phn$\pm$\phn0.191    \\
\enddata
\label{photometry}
\end{deluxetable}

\begin{deluxetable}{lll}
\tablecaption{Results from CHORIZOS Fitting for Sandage 96.}
\tablewidth{0pt}
\tablehead{  & \colhead{Young model} & \colhead{Old model}}
\startdata
Age (Myr)                                  & \phs13.6\phn\phn\phn$\pm$\phn0.2   & \phs28.8\phn\phn\phn$\pm$\phn2.2   \\
\ebv\ (mag)                                &    \phs\phn0.172\phn$\pm$\phn0.022 &    \phs\phn0.277\phn$\pm$\phn0.021 \\
$M_V$                                      &     $-$10.08\phn\phn$\pm$\phn0.09  &     $-$10.39\phn\phn$\pm$\phn0.07  \\
Mass ($10^3\,M_\odot$)\tablenotemark{a}    & \phs24.0\phn\phn\phn$\pm$\phn1.1   & \phs57.5\phn\phn\phn$\pm$\phn2.6   \\
$\chi^2$ per degree of freedom             &     \phs\phn0.28                   & \phs\phn1.74                       \\
K+M stars, types I and II\tablenotemark{b} &           \phs12                   & \phs50                             \\
B stars, types I and II\tablenotemark{b}   &      \phs\phn2.5                   & \phs17                             \\
B stars, type III\tablenotemark{b}         &           \phs24                   & \phs89                             \\
\enddata
\tablenotetext{a}{Assuming a Salpeter IMF between $1\,M_\odot$ and
$100\,M_\odot$.}
\tablenotetext{b}{Predicted number of stars, from Starburst99.}
\label{results}
\end{deluxetable}

\begin{figure}
\begin{center}
\centerline{\includegraphics*[width=0.48\linewidth]{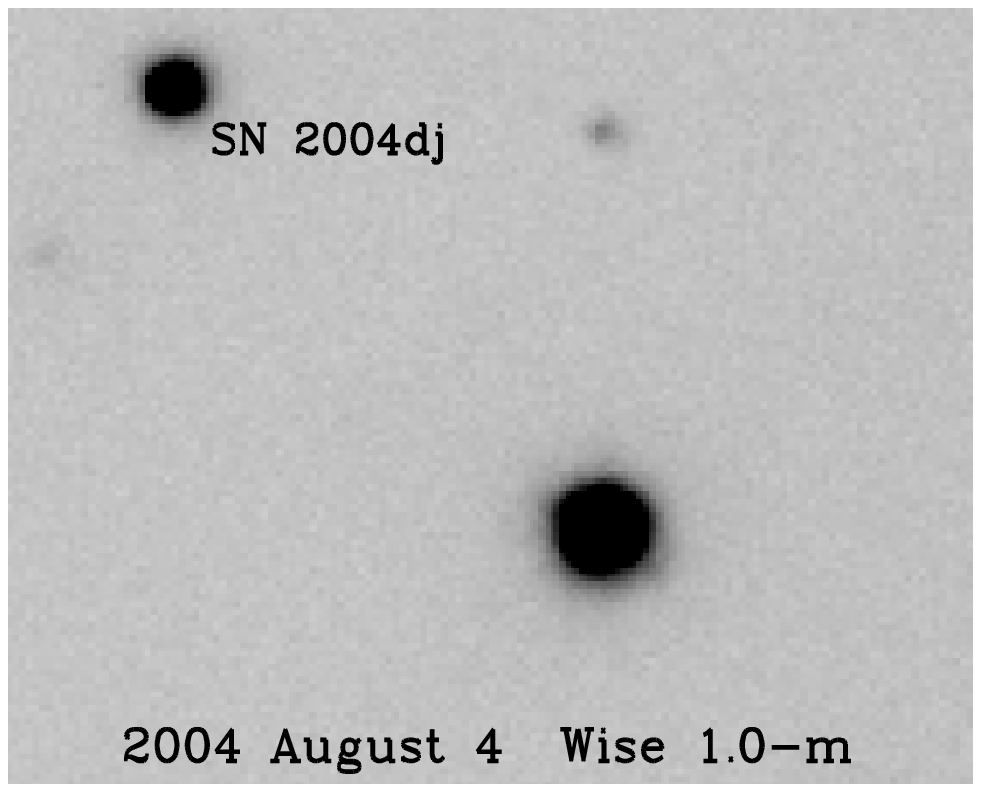} \
            \includegraphics*[width=0.48\linewidth]{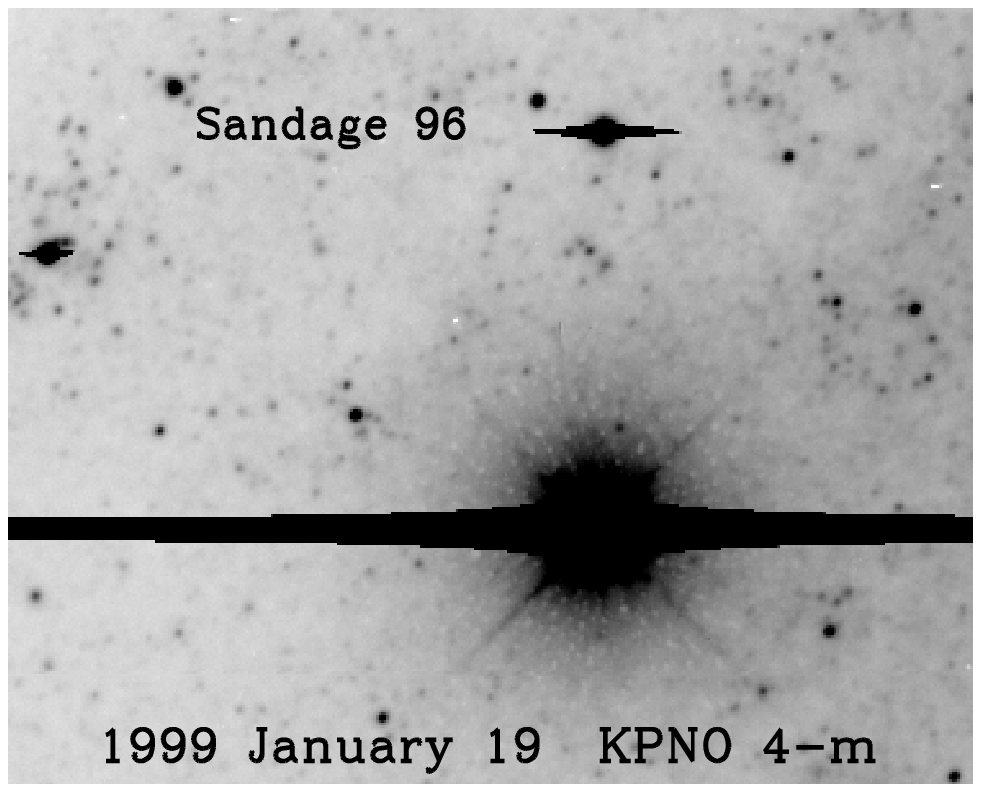}}
\end{center}
\figcaption{CCD images of the site of SN 2004dj taken on 2004 August 4 
(left; Wise Observatory 1.0-m, $V$ band, 60-s exposure) and 1999 January 19
(right; Kitt Peak 4-m, $I$ band, 600-s exposure). North is at the top
and east is on the left, and the images are $96\arcsec$ wide. The supernova
coincides with Sandage~96 to within $0\farcs07$.}
\end{figure}

\clearpage

\begin{figure}
\centerline{\includegraphics*[width=\linewidth, bb=28 28 566 514]{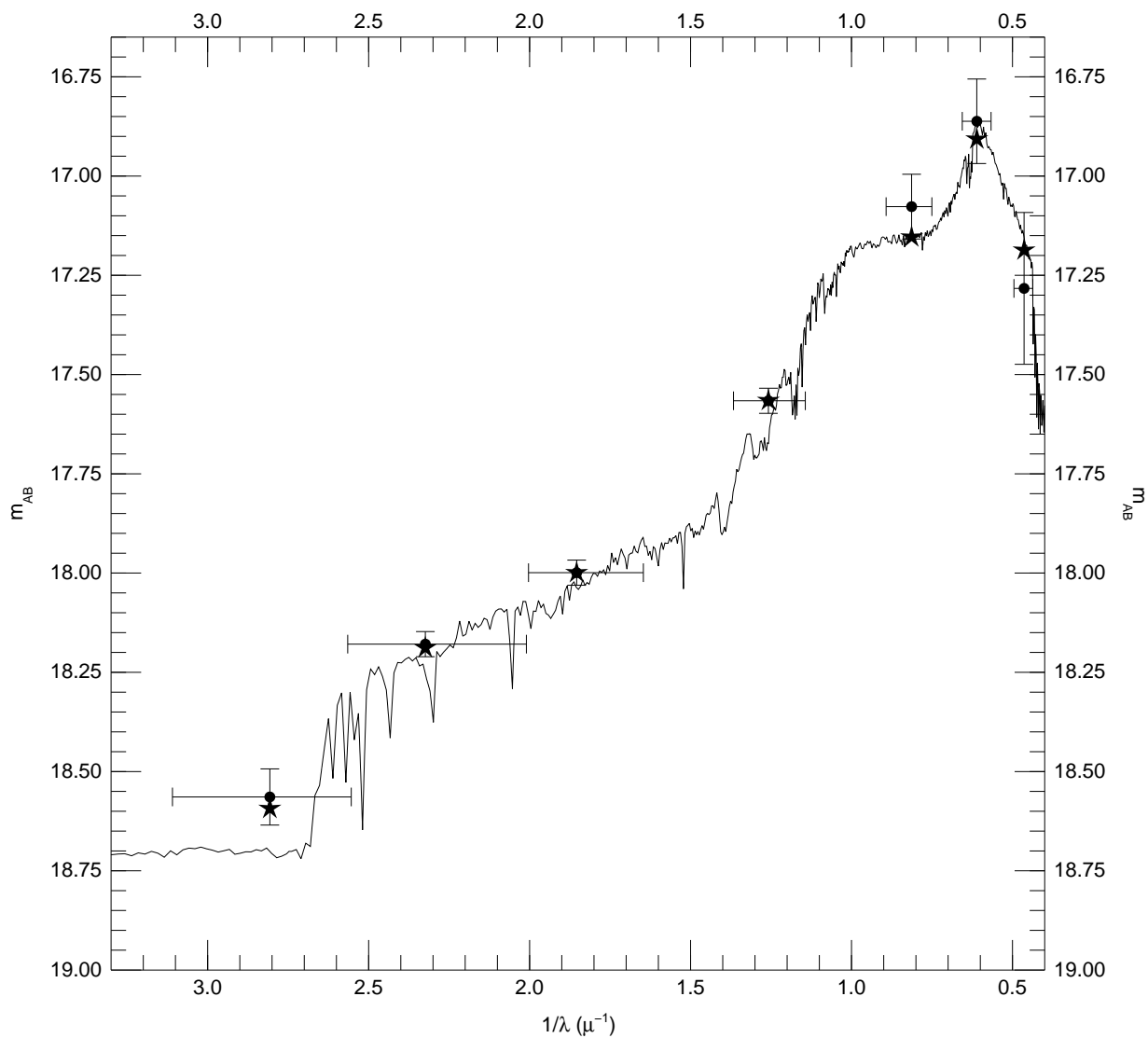}}
\caption{SED for the best fit (13.6 Myr solution) to the optical-NIR photometry of
Sandage~96, based on Starburst99 models of integrated stellar populations. 
The photometry itself is shown by the symbols with error bars (vertical ones
for uncertainties and horizontal ones for the  approximate wavelength coverage
of each filter). Star symbols indicate the calculated magnitude of the model
SED for each filter.}
\label{spectra}
\end{figure}

\begin{figure}
\centerline{\includegraphics*[width=0.8\linewidth, bb=28 28 566
538]{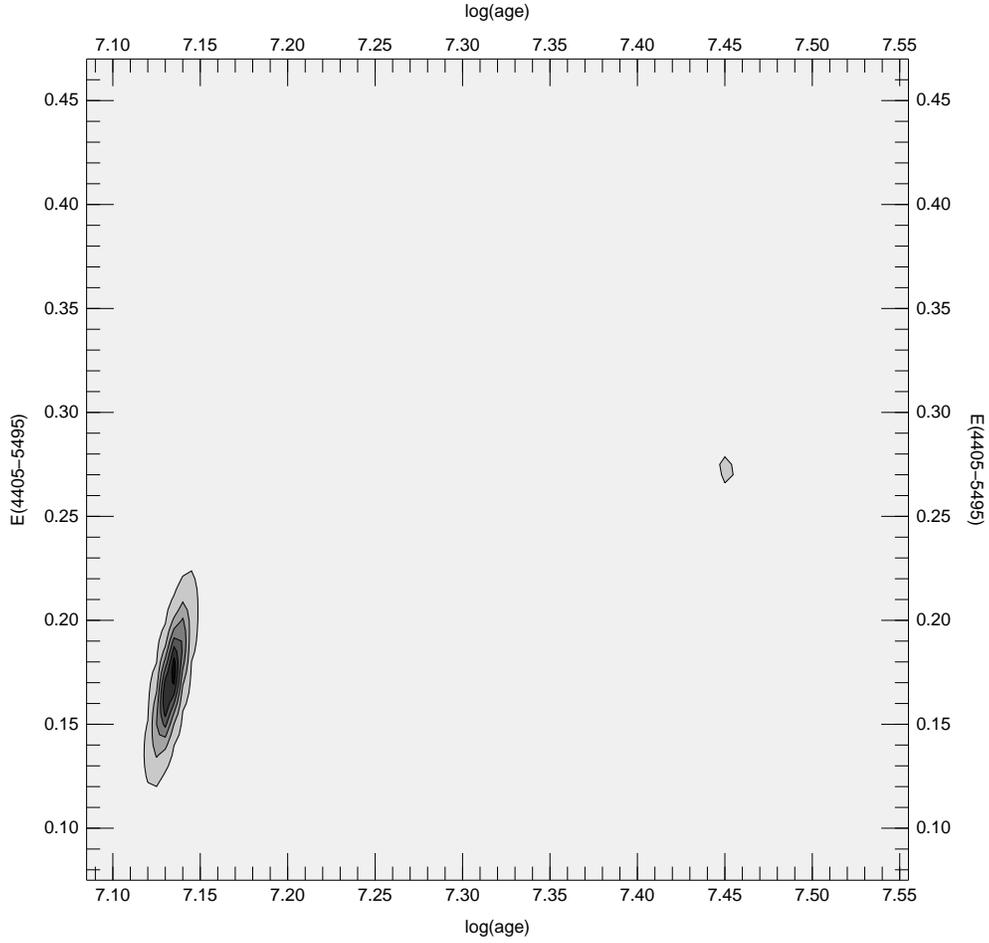}}
\caption{Likelihood contour plot in reddening vs.\ age for the CHORIZOS fit
using Starburst99 models. Age is expressed in years. The favored solution has a
cluster age of 13.6 Myr and a reddening of 0.172~mag.}
\label{likelihood}
\end{figure}

\end{document}